# Enhanced Permittivity in Wurtzite ScAlN through Nanoscale Sc Clustering

James L Hart[1]*, Andrew C Lang[2]**, Matthew T Hardy[2], Saikat Mukhopadhyay[2], Vikrant J Gokhale[2], James G. Champlain[2], Bethany M. Hudak[2], Gabriel Giribaldi[3], Luca Colombo[3], Matteo Rinaldi[3], Brian P Downey[2]

1. Nova Research Inc, 1900 Elkin St # 230, Alexandria, VA
2. U.S. Naval Research Laboratory, 4555 Overlook Ave SW, Washington, DC
3. Institute for NanoSystems Innovation (NanoSI), Northeastern University, Boston MA

*Corresponding author email: james.hart243.ctr@us.navy.mil
**Corresponding author email: andrew.c.lang4.civ@us.navy.mil

ScN alloyed AlN ($Sc_xAl_{1-x}N$, ScAlN) is a wurtzite semiconductor with attractive ferroelectric, dielectric, piezoelectric, and optical properties. Here, we show that ScAlN films (with $x$ spanning 0.18 to 0.36) contain nanoscale Sc-rich clusters which maintain the wurtzite crystal structure. While both molecular beam epitaxy (MBE) and sputter deposited $Sc_{0.3}Al_{0.7}N$ films show Sc clustering, the degree of clustering is significantly stronger for the MBE-grown film, offering an explanation for some of the discrepancies between MBE-grown and sputtered films reported in the literature. Moreover, the MBE-grown $Sc_{0.3}Al_{0.7}N$ film exhibits a dispersive and anomalously large dielectric permittivity, roughly double that of sputtered $Sc_{0.3}Al_{0.7}N$. We attribute this result to the Sc-rich clusters locally reaching $x \sim 0.5$ and approaching the predicted ferroelectric-to-paraelectric phase transition, resulting in a giant (local) enhancement in permittivity. The Sc-rich clusters should similarly affect the piezoelectric, optical, and ferroelectric responses, suggesting cluster-engineering as a means to tailor ScAlN's functional properties.

$Sc_xAl_{1-x}N$ – the first discovered wurtzite ferroelectric – is an emerging nitride semiconductor with potential applications including neuromorphic computing, high-power RF electronics, acousto-electronics for RF signal processing, and electro-optics[1-5]. Despite this promise, ScAlN remains an immature material system with important questions unresolved. In particular, there are large discrepancies in the literature regarding basic properties of ScAlN, e.g. composition versus lattice constant[6-8], film leakage current[9-11], dielectric constant[4,12], etc. There are also unexplained differences between ScAlN grown by molecular beam epitaxy (MBE) and sputter deposition[12]. Resolving these issues is necessary for the continued advancement of ScAlN and will require more in-depth materials characterization and defect analysis.

The ScN-fraction, $x$, is an important parameter which affects essentially all of ScAlN's functional properties[6-8,13]. For certain applications, *e.g.* piezoelectric filters[14], the figure of merit scales with $x$, thus there is a desire to grow high ScN-fraction films[12,15,16]. At very high ScN-fraction ($x \sim 0.6$), theoretical studies predict a transition from the ferroelectric wurtzite phase to its high symmetry paraelectric counterpart, the so-called layered hexagonal phase[17,18]. Materials close to phase transitions often display giant response functions as well as electric-field tunability[19,20], making this predicted transition of theoretical and technological importance. However, ScAlN films with $x > 0.3$ typically display increased structural disorder and contain secondary phases, specifically, the cubic rock salt phase[21,22]. The structural quality is generally worse for films with $x \sim 0.4$[12,15,23], and films with $x \sim 0.5$ are majority rock salt[18].

For alloyed systems, the presence of chemical short range order (SRO) can dramatically influence material properties, with prominent examples including high entropy alloys[24], relaxor ferroelectrics[25], and semiconductor alloys[26]. For wurtzite ScAlN, several theoretical studies predict a positive enthalpy of mixing, which should drive Sc clustering through spinodal decomposition[17,18]. Experimentally, however, there are conflicting reports of chemical SRO in ScAlN, with several observations complicated by the presence of cubic inclusions[18,22,27-30]. Resolving this question is important in two regards. First, if the nature and degree of SRO differs between samples, this effect may explain some of the inconsistencies in the field. Second, the concept of SRO engineering is gaining traction across the materials science community and may prove useful in controlling the properties of ScAlN.

Here, through a comprehensive set of experimental and theoretical methods, we show that nanoscale Sc clustering is a general phenomenon in ScAlN, with evidence of clustering observed in all studied films: MBE-grown films with $x$ = 0.18, 0.3, and 0.36, and a sputter-deposited film with $x$ = 0.3. By comparing the $x$ = 0.3 MBE and sputtered films, we show that the degree of clustering is significantly stronger for the MBE-grown film. As reported previously[12], this $Sc_{0.3}Al_{0.7}N$ MBE-grown film shows an anomalously large and dispersive dielectric permittivity ($k \sim 60$ at 100 Hz to $k \sim 30$ at 1 GHz), while the sputtered film shows a frequency-independent permittivity of $k \sim 17$. We argue that the enhanced permittivity of the MBE-grown film is a direct result of the Sc-rich clusters with $x \sim 0.5$ locally approaching the ferroelectric-to-paraelectric phase transition. Hence, engineered Sc clustering may offer access to the desirable properties of high ScN-fraction ScAlN while retaining the phase purity and high crystal quality of films with $x \leq 0.3$. We note, however, that in our samples, Sc clustering is correlated with increased DC leakage and dielectric loss, which are detrimental for some device applications. Overall, our work highlights the presence and importance of Sc clustering in ScAlN and calls for future research to better control the local composition during growth.

**Results:**

*Theoretical Investigation of Sc Clustering:*

To begin, we study the energetics of Sc clustering through a multi-scale theoretical analysis. We performed density functional theory (DFT) calculations of 96 atom ScAlN supercells (Supplementary Note 1)[31]. In total, we computed the ground state energy of 141 ScAlN structures, with concentrations varying from 0% to 50% Sc. Figure 1a plots the mixing enthalpy for all the supercells, with $H_{mix}$ defined as

$$H_{mix} = E - xE_{ScN} - (1-x)E_{AlN} \qquad \text{Equation 1}$$

where $E_{ScN}$ is the energy of rock-salt ScN, and $E_{AlN}$ is the energy of wurtzite AlN. We find that the mixing enthalpy of wurtzite ScAlN is positive, with negative $d^2H/dx^2$ across the studied composition range. This result is consistent with a number of other works[17,18] and indicates an energetic instability towards cluster formation. Inspecting the supercells with 50% Sc, the highest energy configuration contains all Sc atoms arranged on alternating basal planes (Fig. 1a, top right schematic). Conversely, the lowest energy configurations contain Sc clusters elongated along the *c*-axis (Fig. 1a, bottom right schematic).

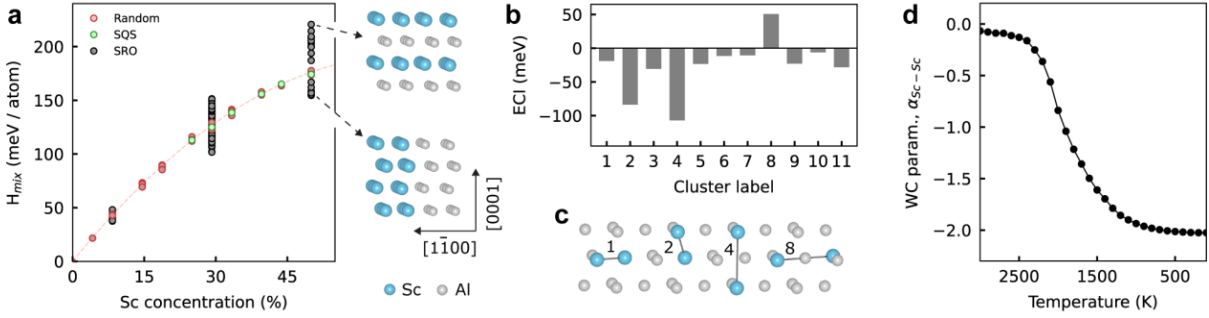

**Figure 1. Theoretical investigation of Sc clustering. a)** DFT calculated mixing enthalpy of wurtzite ScAlN as a function of Sc concentration. Red line is a 2$^{nd}$ order polynomial fit to the special quasi-random (SQS) and random structures. The 'SRO' datapoints at $x$ = 0.1, 0.3, and 0.5 correspond to structures with specifically tailored clustered and anti-clustered configurations, see Supplementary. The atomic structures for two supercells with $x$ = 0.5 are shown, with the N sublattice excluded for clarity. **b)** Cluster energies for all of the two-body clusters included in our CE model. **c)** Schematic of several two-body clusters, with the N sublattice excluded for clarity. The cluster numbers correspond to the labels in panel **b**. **d)** MC simulation for Sc$_{0.3}$Al$_{0.7}$N, plotting the Warren-Cowley parameter α as a function of temperature upon cooling.

Next, we analyzed the DFT computed energies with a cluster expansion (CE) model[32] (Supplementary Note 2):

$$E_{CE}(\sigma) = J_0 + \sum_i J_1 \sigma_i + \sum_i \sum_j J_{ij} \sigma_i \sigma_j + \sum_i \sum_j \sum_k J_{ijk} \sigma_i \sigma_j \sigma_k \qquad \text{Equation 2}$$

where σ describes the Sc configuration of the supercell. The index *i* includes all metal sites, with σ = 0 for Al and σ = 1 for Sc. The *J*s are the effective cluster interaction (ECI) energies which are fit to the DFT data. Importantly, $J_{ij}$ is the two-body interaction energy between Sc atoms located at $R_i$ and $R_j$. Figure 1b shows all of the two-body cluster energies, and Fig. 1c graphically represents the most important two-body clusters. Excluding cluster #8, all of the of the two-body ECIs are negative, indicating an energetic driving force towards Sc clustering. In particular, clusters which are primarily out-of-plane are energetically favorable (see #2 and 4). The three-body ECIs are given in the Supplementary and are all less than 30 meV.

To understand how the microscopic ECIs dictate larger scale clustering, we performed Monte Carlo (MC) simulations using a 26,208 atom Sc$_{0.3}$Al$_{0.7}$N supercell (Supplementary Note 3). Our MC simulations are constrained to keep the crystal structure in the wurtzite phase; only the configuration of Sc and Al atoms was allowed to vary. To track the degree of Sc clustering as a function of temperature, we calculate the Warren-Cowley parameter[33],

$$\alpha_{Sc-Sc} = 1 - \frac{CN_{Sc-Sc}/CN_{Sc-M}}{x} \qquad \text{Equation 2}$$

where $CN_{Sc-Sc}$ is the average number of Sc-Sc bonds per Sc atom, and $CN_{Sc-M}$ is the total number of Sc-M (M = metal) bonds per Sc atom. For this calculation, we consider the nearest-neighbor metal-metal bonds with $R \sim 3$ Å and $CN_{Sc-M} = 12$. This shell consists of 6 in-plane bonds and 6 out-of-plane bonds, corresponding to clusters #1 and 2 in Fig 1c, respectively. Negative values of α indicate clustering, and positive values indicate anti-clustering. The simulations show clustering for temperatures < 2000 K. Hence, we find strong theoretical evidence that wurtzite ScAlN is susceptible to spinodal decomposition and the formation of Sc-rich clusters.

*Experimental Confirmation of Sc Clustering:*

Experimentally, determination of SRO is difficult: the signatures of SRO are weak, subject to multiple interpretations, and prone to artifacts[34,35]. To address this challenge, we study SRO through several independent and complementary experimental probes, including reciprocal space diffuse scattering, real-space elemental mapping, and extended fine structure analysis. We first consider the $x = 0.3$ (nominal) ScAlN film grown via MBE (growth details in Methods).

We measure the reciprocal-space diffuse scattering via two methods: selected area electron diffraction (SAED) and Fourier analysis of atomic-resolution high angle annular dark field scanning transmission electron microscopy HAADF-STEM images. In both cases, we observe significant diffuse scattering around the wurtzite Bragg spots (Fig. 2a,b). This is the expected signal for Sc clustering (Supplementary Figure 1)[36,37]. The diffuse scattering is strongest along the $\langle 1\bar{1}01 \rangle$ directions, which indicates that in real-space the clusters prefer the pyramidal planes. This result is reproduced in our MC simulations (Supplementary Figure 1) and reflects the fact that the two-body Sc-Sc energetics are strongly anisotropic (Fig. 1b).

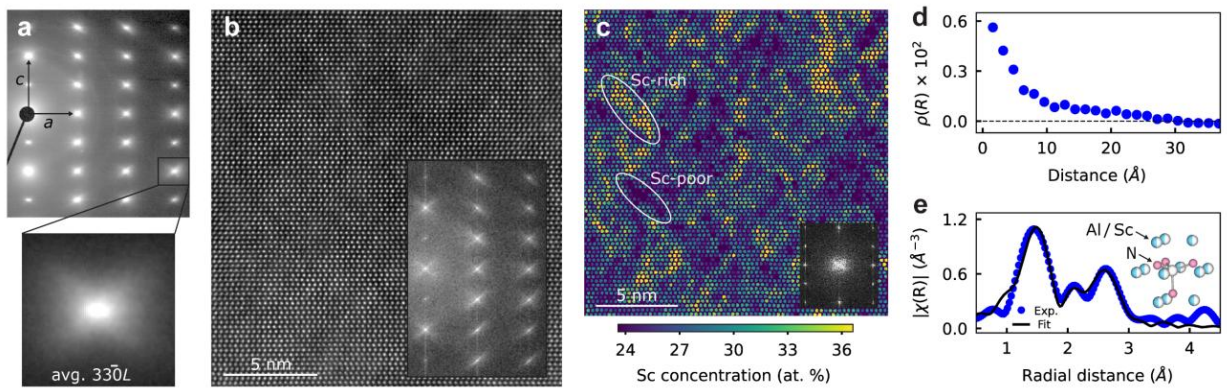

**Figure 2. Experimental observation of clustering in MBE-grown Sc$_{0.3}$Al$_{0.7}$N. a)** Electron diffraction pattern of MBE-grown Sc$_{0.3}$Al$_{0.7}$N. Data collected down the [11$\bar{2}$0] zone axis, and presented on a log scale. The zoom-in presents the averaged intensity of all $3\bar{3}0L$ spots, with $|L| <$ 3. **b)** HAADF-STEM image of the same film and zone-axis. The displayed image is cropped from a larger image (Supplementary Figure 2). The inset shows the Fourier transform (FT) of the larger image, after application of a 2D Hanning window. **c)** STEM-EELS derived map of the Sc concentration for each atomic column. The inset shows the FT of the raw STEM-EELS intensity map. **d)** 2D auto-correlation of the Sc concentration calculated from the data in panel **c** and

Equation 3. **e)** Al *K*-edge extended fine structure in *R*-space. The inset shows the local structure around Al.

To visualize the Sc clusters in real-space, we next performed STEM electron energy-loss spectroscopy (EELS). We processed the EELS data to determine the Sc concentration for each atomic-column, as presented in Fig. 2c (Supplementary Note 4). Visually, both Sc-rich clusters and Sc-poor clusters are observed in the EELS map. To quantify the clustering, we calculate the 2D autocorrelation of the Sc concentration,

$$\rho(R) = <(x(r) - \mu)(x(r-R) - \mu)> \qquad \text{Equation 3}$$

where *R* is the separation between atomic columns, $<...>$ denotes the spatial average, $x(r)$ is the Sc composition at position *r*, and μ is the signal mean. A positive value of ρ indicates that Sc-rich atomic columns are grouped together, as expected for clustering. Experimentally, we observe a positive and finite ρ out to nearly 3 nm (Fig. 2d), confirming the presence of Sc clusters and providing an estimate of their size. The EELS data also shows the cluster anisotropy. As highlighted with the white ovals in Fig. 2c, both Sc-rich and Sc-poor clusters are observed along pyramidal planes. Additionally, the FT of the raw EELS map (see inset) shows the same diffuse scattering as the SAED and HAADF-STEM data.

To determine the local composition of the Sc-rich clusters, we next performed Al *K*-edge extended energy-loss fine structure analysis (EXELFS)[38,39]. By fitting the Al *K*-edge extended fine structure, we are able to extract the average local environment of Al. Specifically, we are interested in the average number of Al-Sc bonds within the nearest-neighbor metal-metal shell. For a random solid solution, the average *local* composition should match the *global* composition, that is, $CN_{Al-Sc} / CN_{Al-M} = x = 0.3$. Thus, a value of $CN_{Al-Sc} \sim 3.6$ is consistent with a random alloy. Conversely, for clustering, we expect an excess of Al-Al bonds, resulting in $CN_{Al-Sc} < 3.6$.

Figure 2d shows the EXELFS data in *R*-space, with the processing and fitting details provided in Supplementary Note 5. The fitted value of $CN_{Al-Sc}$ is $2.2 \pm 0.6$, confirming Sc clustering. From this result, simple algebra yields the local Sc environment, which is $CN_{Sc-Sc} = 6.7 \pm 1.4$ (Supplementary Note 6). Thus, despite a global composition of 30% Sc, the average *local* Sc environment is $56 \pm 12\%$ Sc. This result shows that the Sc-rich clusters approach the critical value of $x \sim 0.6$. Our real-space EELS measurements (Fig. 2c) are consistent with the Sc-rich clusters having a composition of >50% Sc after accounting for through-thickness averaging (Supplementary Note 7).

For bulk ScAlN with $x \sim 0.56$, calculations indicate that the cubic rock-salt phase is the ground state structure[17,18]. While the SAED data shows faint cubic reflections ($10^3$ times weaker that the wurtzite spots), there is no evidence of cubic inclusions within our atomic resolution imaging or EELS data. Moreover, the HAADF-STEM Fourier transform (Fig. 2b) shows no cubic spots, and the EXELFS fit is consistent with phase-pure wurtzite. We conclude that the Sc-rich clusters exist within the wurtzite crystal structure and are not associated with cubic inclusions. The Sc-rich clusters could be stabilized in the wurtzite phase owing to their nanoscale morphology. It is also possible that the Sc-rich clusters are meta-stable and kinetically trapped during growth.

We next determine whether the Sc-rich clusters possess a reduced polar distortion, as would be expected if the clusters are near the ferroelectric-to-paraelectric transition. The polar distortion is quantified by the internal parameter $u$, which defines the separation of the M- and N-layers as a fraction of the $c$ lattice parameter: $u = 0.381$ for AlN and $u = 0.5$ for the theoretical layered hexagonal phase (Fig. 3a)[40]. To locally measure $u$, we use multi-slice electron ptychography (MEP), as presented in Figure 3b. The studied region is entirely monodomain wurtzite. However, image quantification (Supplementary Note 8) reveals variation in the local polar distortion, ranging from $u = 0.437$ down to $0.383$ (Fig. 3c). The spatial evolution of the internal parameter $u$ and the Sc composition $x$ are shown in Figures 3d and 3e, respectively. A correlation between $u$ and $x$ is visually evident, and the calculated Pearson correlation coefficient is 0.47. This result confirms that the Sc-rich clusters locally suppress the polar distortion. According to DFT simulations[40,41], as the Sc composition increases from $x = 0$ to 0.5, $u$ gradually increases from 0.381 to ~0.415. Above $x \sim 0.5$, $u$ rapidly increases through the phase transition, reaching $u = 0.5$ by $x \sim 0.6$. Hence, the clusters shown in Fig. 3d with $u > 0.415$ are indicative of local regions at the cusp of the ferroelectric-to-paraelectric phase transition.

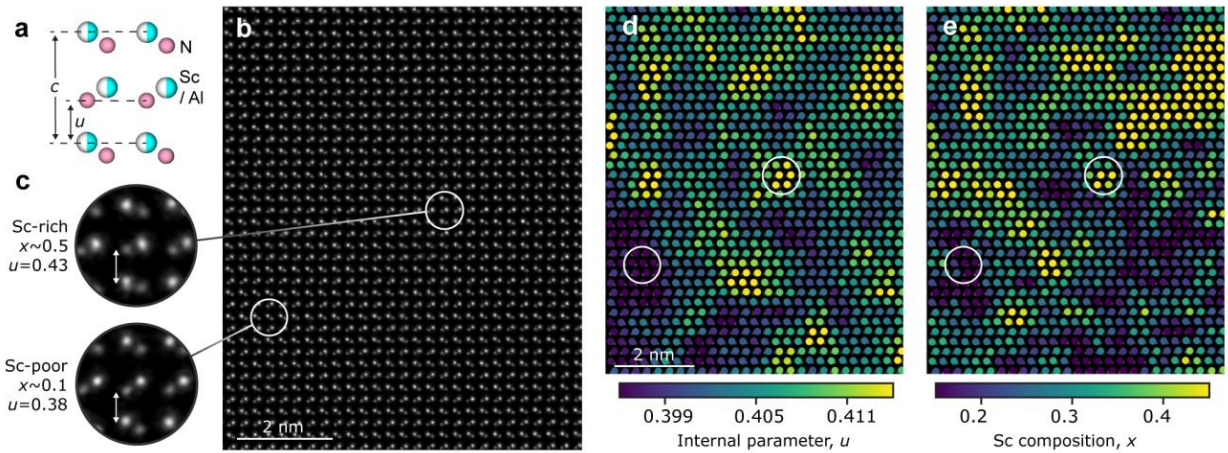

**Figure 3. Correlation of local Sc composition and polarization in MBE-grown $Sc_{0.3}Al_{0.7}N$. a)** Schematic showing the internal parameter $u$ which defines the AlN wurtzite phase ($u = 0.381$) and the layered hexagonal phase ($u = 0.5$). Note that $u$ is defined as a fraction of the lattice constant $c$. **b)** Multi-slice electron ptychography reconstruction of the MBE-grown $Sc_{0.3}Al_{0.7}N$ film, showing the phase of a single slice taken from the middle of the specimen. **c)** Zoomed-in regions of the MEP dataset showing variation in $u$ and $x$. Note that the $x \sim 0.1$ region is flipped vertically to match the $x \sim 0.5$ region. Maps of $u$ and $x$ are provided in **d** and **e**, respectively, obtained by quantification of the dataset shown in **b**. The $u$ and $x$ maps are positively correlated, with a Pearson coefficient of 0.47. To aid visualization, both maps are gaussian smoothed with a full width at half maximum of 1.5 unit cells. The white circles mark the two regions highlighted in panel **c**.

Having established the presence of Sc-rich clusters with $x \sim 0.5$ in our MBE-grown film, we next examine the sputter-deposited $Sc_{0.3}Al_{0.7}N$ film. While diffuse scattering is observed for the sputtered sample (Fig. 4a), the signal is weaker and more isotropic compared to the MBE-grown film. This is highlighted in Fig 4b, which plots the azimuthal dependence of diffuse scattering around the central spot. From atomic-resolution STEM-EELS measurements of the sputtered sample, we calculate the Sc auto-correlation function, which is plotted in Fig. 4c. While the data indicates Sc clustering, the correlation decays to zero within ~1.5 nm. For comparison, the MBE-grown film shows a finite correlation out to ~3 nm. This suggests a smaller cluster size in the sputtered film. EXELFS analysis on the sputtered sample yields an average Sc environment of 40 ± 14% Sc, in contrast to 56 ± 12% Sc for the MBE-grown film (Supplementary Figure 3). Hence, while all of our measurements show that the sputtered ScAlN film possesses Sc clustering, the clusters are smaller, more isotropic, and less Sc-rich relative to the MBE film. This result is likely related to the lower growth temperature and higher deposition rate of sputtering, which would restrict the kinetics of cluster formation (Methods).

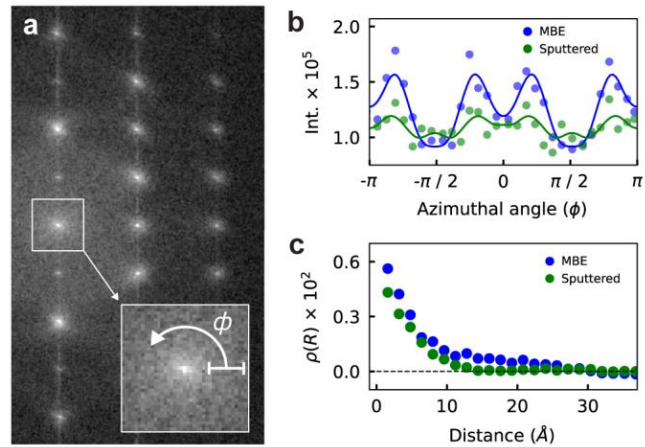

**Figure 4. Comparison of clustering in sputtered vs MBE-grown $Sc_{0.3}Al_{0.7}N$. a)** Fourier transform of a HAADF-STEM image of the sputtered $Sc_{0.3}Al_{0.7}N$ film. The STEM image is shown in Supplementary Figure 3. The inset highlights the central spot. **b)** Azimuthal dependence of the diffuse scattering from $Q = 0.4 – 1$ nm$^{-1}$ (integration limits shown in the panel **a** inset). The MBE data corresponds to the FT shown in the Fig. 2b inset. The solid lines are a Savitzky-Golay filter after symmetrization of the raw data, which enforces $I(\phi) = I(-\phi) = I(\phi + \pi)$. **c)** Auto-correlation of the Sc concentration for the sputtered $Sc_{0.3}Al_{0.7}N$ film. The STEM-EELS data is presented in Supplementary Figure 3. The MBE data is the same as Fig 2d.

In addition to the MBE and sputtered $Sc_{0.3}Al_{0.7}N$ films, we also find evidence of Sc clustering in MBE-grown $Sc_{0.18}Al_{0.82}N$ and $Sc_{0.36}Al_{0.64}N$ (Methods, Supplementary Figure 6). In agreement, our MC simulations indicate that ScAlN is susceptible to cluster formation across a broad range of compositions (Supplementary Figure 7). Thus, while the degree of clustering is dependent upon the growth method (and likely dependent upon the specific growth parameters), Sc clustering appears to be a general phenomenon in ScAlN.

*Clustering Effects on Material Properties:*

In a prior work, we measured the broad-band complex permittivity of the MBE and sputtered $Sc_{0.3}Al_{0.7}N$ films[12]. While the sputtered film shows a frequency-independent real permittivity of $k = 17$, the MBE film possesses a much larger and frequency-dispersive permittivity, ranging from $k = 60$ at 100 Hz to $k = 30$ at 1 GHz. Based on numerous materials characterization methods – including X-ray diffraction, energy-dispersive X-ray spectroscopy, atomic-force microscopy, basic STEM imaging, SAED, and secondary ion mass spectroscopy – we found no substantive differences between the sputtered and MBE-grown films[12]. The most striking difference between samples is the degree of Sc clustering, as reported in this article.

Figure 5 depicts our argument that Sc clustering leads to enhanced permittivity. The black curve shows the permittivity as a function of Sc composition, as determined from the Landau–Devonshire model in ref. 42. The permittivity diverges as ScAlN approaches the ferroelectric-to-paraelectric transition at $x \sim 0.6$. For a clustered sample, macroscopic permittivity measurements will be influenced by both the Sc-poor clusters and the Sc-rich clusters. Given the non-linear curve, the increased permittivity from the Sc-rich clusters will more than offset the decreased permittivity in the Sc-poor clusters. As a first-order approximation, we use our EXELFS results and the data from ref. 42 to estimate the ensemble permittivity of our MBE film, which yields $k = 44$ (Supplementary Note 6). This value is within the bounds of our experimental data. The nanoscale Sc-rich clusters can also explain the dispersive properties of the MBE-grown film, as dispersion is common for systems near a ferroelectric-to-paraelectric transition[19,20]. For the sputtered film, our EXELFS data yields a predicted $k$ of 20, in reasonable agreement with the experimental value of $k = 17$.

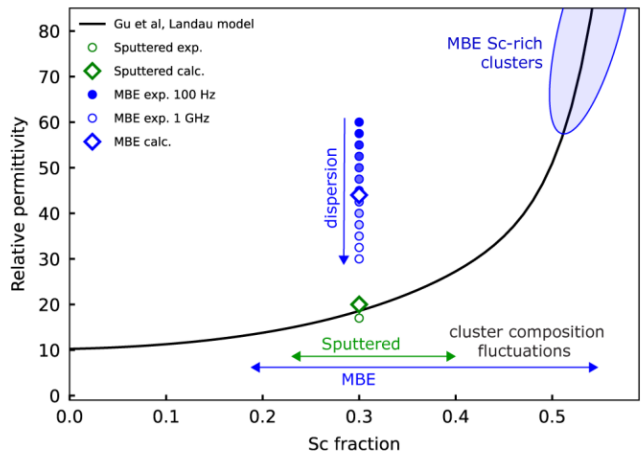

**Figure 5. Comparison of growth method, Sc clustering, and dielectric permittivity.** Real part of the permittivity as a function of Sc fraction. The Landau-Devonshire model in ref. 42 is shown in black. The MBE and sputtered $Sc_{0.3}Al_{0.7}N$ films considered in this work are shown in blue and green, respectively[12]. For these two samples, the 'cluster composition fluctuations' are determined from EXELFS analysis. The 'calculated' datapoints are based on the data in ref. 42 and the EXELFS analysis (Supplementary Note 6).

**Discussion:**

As the growth of wurtzite ScAlN with $x > 0.5$ is not currently possible, Sc clustering offers an alternative means to access the very high permittivity of ScAlN with large ScN-fraction. Sc clustering should similarly affect ScAlN's piezoelectric constant, the ferroelectric coercive field, and the optical non-linearity. Moreover, clustering may lead to enhanced electric-field tunability, which is often observed for systems near a ferroelectric-to-paraelectric transition and is useful for applications such as tunable filters and antennas[19,20]. Thus, there is potential to engineer the Sc clusters for specific applications. However, our $Sc_{0.3}Al_{0.7}N$ MBE-grown film shows significantly higher DC leakage and dielectric loss relative to the sputtered film. The DC leakage could be due to the Sc-rich clusters locally reducing the bandgap and forming a percolative network[43]. It is also possible that the increased leakage in the MBE film is due to an entirely separate defect, *e.g.* N vacancies[10]. Regardless, for the samples studied here, the MBE film with enhanced clustering is too lossy for RF device applications. Minimizing the degree of Sc clustering may thus be preferrable if the ScAlN films are to be used as dielectric or piezoelectric layers.

In conclusion, we report the presence of Sc clustering in wurtzite ScAlN. The clusters are on the nanometer scale, prefer the pyramidal planes, can reach local compositions exceeding 50% Sc, and locally suppress the polar distortion. For sufficiently strong clustering, we argue that Sc-rich regions yield a giant (local) enhancement in permittivity. Looking forward, control of Sc clustering could enable access to the rich physics of materials near phase transitions, potentially offering improved device performance and novel functionality. However, more work is needed to understand how growth parameters influence the degree of clustering and the resulting trade space in material performance.

**Methods:**

The MBE ScAlN samples were grown using a high temperature effusion cell to provide Sc flux, a dual filament effusion cell to provide Al flux, and an RF nitrogen plasma source to provide active nitrogen flux. The Sc and Al fluxes were determined using calibration samples grown at similar compositions, using X-ray photoelectron spectroscopy[21] to determine the ScN fraction, and X-ray reflectivity to determine the layer thickness. The 80 nm thick $Sc_{0.18}Al_{0.82}N$ sample was grown N-rich with a nominal III/V ratio of 0.8 on a 100 nm GaN buffer layer on a N-polar GaN substrate at a nominal substrate temperature of 600 °C[44]. The 150 nm $Sc_{0.30}Al_{0.70}N$ and $Sc_{0.36}Al_{0.64}N$ samples were grown at a substrate temperature of 400 °C, nominal III/V ratio of 0.9, and growth rate of 0.65 Å/s. The $Sc_{0.36}Al_{0.64}N$ sample contained a grade from $Sc_{0.30}Al_{0.70}N \rightarrow Sc_{0.36}Al_{0.64}N$ over the first 25 nm of the film, followed by 125 nm of $Sc_{0.36}Al_{0.64}N$ to prevent formation of rock-salt grains at the lower interface[45], while the $Sc_{0.30}Al_{0.70}N$ film was a single composition over the 150 nm thick film. The composition of the MBE-grown $Sc_{0.3}Al_{0.7}N$ was checked via STEM energy dispersive spectroscopy (EDS) and secondary ion mass spectroscopy (SIMS), both giving 29% Sc.

The MBE $Sc_{0.30}Al_{0.70}N$ and $Sc_{0.36}Al_{0.64}N$ samples were grown continuously on 30 nm AlN / 50 nm NbN templates on 3 inch 4H-SiC wafers. The first half of the AlN layer was grown N-rich at 800 °C to avoid reaction of free Al with the NbN layer, and the second half was grown metal-rich to improve the surface morphology. The NbN layer was grown at an estimated temperature of 650 °C [46] using Nb flux from an electron beam evaporator. A nominally identical NbN layer was used as a template for the sputtered $Sc_{0.30}Al_{0.70}N$ film.

The sputtered film was deposited at 300 ºC with deposition conditions as described in ref. 47. The composition of the sputtered $Sc_{0.3}Al_{0.7}N$ film was measured with STEM-EDS and SIMS, giving 28% Sc and 30% Sc, respectively.

Density functional theory (DFT) calculations were performed using the GPAW code within the projector augmented wave formalism[31]. We use a plane-wave basis set with a cut-off energy of 600 eV, a Γ-centered 2×2×2 $k$-point mesh, and the PBE exchange-correlation functional. All ScAlN supercells studied with DFT contain 96 atoms and are approximately 1×1×1 nm$^3$ orthogonal cells. We relax the system volume, the $c$ to $a$ ratio, and the atomic coordinates until all forces are below 10 meV / Å.

The cluster expansion (CE) analysis and Monte Carlo (MC) simulations were completed with custom python scripts. In our CE formulation, $J_0$ is the energy of the AlN supercell, $J_1$ is the energy to add an isolated Sc atom to the supercell, $J_{ij}$ is the two-body interaction energy between Sc atoms located at $\boldsymbol{R}_i$ and $\boldsymbol{R}_j$, and $J_{ijk}$ is the three-body term. For the MC simulations, each step is accepted with probability $P = \exp(\Delta E/kT)$, where ΔE is the change in system energy for the proposed MC step (determined via the CE effective cluster interaction energies), $k$ is Boltzman's constant, and $T$ is the temperature. The electron diffraction simulations (of the MC supercells) were performed with the abtem package[48].

Electron microscopy specimens were prepared using a ThermoFisher Helios G3 Focused Ion Beam, with the last thinning step performed with 5 kV Ga ions. Final cleaning was then performed by argon ion polishing at 2 eV and then 1 eV using a Fischione 1040 Nanomill.

Selected area electron diffraction (SAED) and extended energy loss fine structure analysis (EXELFS) measurements were made using a JEOL 200F microscope operated at 200 kV. The EXELFS measurements were averaged over a large area of the film (~100 x 500 nm$^2$), and the data was processed using the Athena and Artemis packages[49].

Atomic resolution scanning transmission electron microscope (STEM) images were collected using a Nion UltraSTEM-200 operating at 200 kV with a convergence semi-angle of 26 mrad. STEM electron energy-loss spectroscopy was performed using a Gatan Enfinium ER spectrometer with a Quantum Detector MerlinEM camera, with a collected energy range of 0 – 600 eV. Note that the energy onsets of the Sc $L$-edge and the N $K$-edge overlap. Thus, determining the Sc concentration requires subtraction of the N $K$-signal, which we achieve through comparison with reference measurements of AlN (while assuming negligible N vacancies) and normalization based on the known sample composition (Supplementary Note 4).

The ptychographic measurements were performed on the Nion STEM using the MerlinEM camera, and the reconstruction was performed using the PtyRAD code[50]. Further details regarding the ptychography acquisition, reconstruction, and processing are provided in Supplementary Note 8.

Since the MBE and sputtered samples were prepared for STEM using the same procedure, thinned to the same approximate thickness, and imaged under the same conditions, the EELS and HAADF data can be compared semi-quantitatively (Fig. 4).

In our 200 kV experiments, we did not observe any electron beam induced Sc clustering (Supplementary Figure 4). We also repeated the SAED and EXELFS measurements at 80 kV

(below the presumptive knock-on damage threshold) on a separate STEM specimen, and we observed the same signatures of clustering (Supplementary Figure 5).

**Acknowledgements:**

This work was funded by the US Naval Research Lab Base Program (JH, AL, and BH) and the Defense Advanced Research Projects Agency, Microsystem Technology Office (DARPAMTO) under the Compact Front-end Filters at the ElEment-level (COFFEE) program (MH, SM, VG, JC, GG, LC, MR, BD).

**Author contributions:**

J.H. and A.L. performed the STEM, SAED, EELS, MEP, and EXELFS experiments and analysis, and B.H assisted with the MEP experiments. J.H. performed the DFT calculations and MC simulations. M.H. grew the MBE samples, and G.G., L.C., and M.R. provided the sputtered sample. V.K. performed the dielectric permittivity measurements. J.H. wrote the manuscript. All authors discussed the results and revised the paper.